\documentclass[11pt,a4paper]{article}
\pdfoutput=1
\usepackage{amssymb} \usepackage{amsmath} \usepackage{graphicx}
\usepackage{epsfig,latexsym}
\begin{document}

\def\bef{\begin{figure}}
\def\eef{\end{figure}}
\newcommand{\ans}{ansatz }
\newcommand{\be}[1]{\begin{equation}\label{#1}}
\newcommand{\beq}{\begin{equation}}
\newcommand{\ee}{\end{equation}}
\newcommand{\beqn}[1]{\begin{eqnarray}\label{#1}}
\newcommand{\eeqn}{\end{eqnarray}}
\newcommand{\bd}{\begin{displaymath}}
\newcommand{\ed}{\end{displaymath}}
\newcommand{\mat}[4]{\left(\begin{array}{cc}{#1}&{#2}\\{#3}&{#4}
\end{array}\right)}
\newcommand{\matr}[9]{\left(\begin{array}{ccc}{#1}&{#2}&{#3}\\
{#4}&{#5}&{#6}\\{#7}&{#8}&{#9}\end{array}\right)}
\newcommand{\matrr}[6]{\left(\begin{array}{cc}{#1}&{#2}\\
{#3}&{#4}\\{#5}&{#6}\end{array}\right)}
\newcommand{\cvb}[3]{#1^{#2}_{#3}}
\def\lsim{\raise0.3ex\hbox{$\;<$\kern-0.75em\raise-1.1ex
e\hbox{$\sim\;$}}}
\def\gsim{\raise0.3ex\hbox{$\;>$\kern-0.75em\raise-1.1ex
\hbox{$\sim\;$}}}
\def\abs#1{\left| #1\right|}
\def\simlt{\mathrel{\lower2.5pt\vbox{\lineskip=0pt\baselineskip=0pt
           \hbox{$<$}\hbox{$\sim$}}}}
\def\simgt{\mathrel{\lower2.5pt\vbox{\lineskip=0pt\baselineskip=0pt
           \hbox{$>$}\hbox{$\sim$}}}}
\def\unity{{\hbox{1\kern-.8mm l}}}
\newcommand{\eps}{\varepsilon}
\def\ep{\epsilon}
\def\ga{\gamma}
\def\Ga{\Gamma}
\def\om{\omega}
\def\omp{{\omega^\prime}}
\def\Om{\Omega}
\def\la{\lambda}
\def\La{\Lambda}
\def\al{\alpha}
\newcommand{\ov}{\overline}
\renewcommand{\to}{\rightarrow}
\renewcommand{\vec}[1]{\mathbf{#1}}
\newcommand{\vect}[1]{\mbox{\boldmath$#1$}}
\def\tm{{\widetilde{m}}}
\def\mcirc{{\stackrel{o}{m}}}
\newcommand{\Dm}{\Delta m}
\newcommand{\dm}{\varepsilon}
\newcommand{\tanb}{\tan\beta}
\newcommand{\nbar}{\tilde{n}}
\newcommand\PM[1]{\begin{pmatrix}#1\end{pmatrix}}
\newcommand{\up}{\uparrow}
\newcommand{\down}{\downarrow}
\def\omE{\omega_{\rm Ter}}
%

\newcommand{\Dsusy}{{susy \hspace{-9.4pt} \slash}\;}
\newcommand{\DCP}{{CP \hspace{-7.4pt} \slash}\;}
\newcommand{\mc}{\mathcal}
\newcommand{\gr}{\mathbf}
\renewcommand{\to}{\rightarrow}
\newcommand{\gtc}{\mathfrak}
\newcommand{\wh}{\widehat}
\newcommand{\br}{\langle}
\newcommand{\kt}{\rangle}


\def\lsim{\mathrel{\mathop  {\hbox{\lower0.5ex\hbox{$\sim$}
\kern-0.8em\lower-0.7ex\hbox{$<$}}}}}
\def\gsim{\mathrel{\mathop  {\hbox{\lower0.5ex\hbox{$\sim$}
\kern-0.8em\lower-0.7ex\hbox{$>$}}}}}

\def\nn{\\  \nonumber}
\def\de{\partial}
\def\brf{{\mathbf f}}
\def\bbf{\bar{\bf f}}
\def\bF{{\bf F}}
\def\bbF{\bar{\bf F}}
\def\bA{{\mathbf A}}
\def\bB{{\mathbf B}}
\def\bG{{\mathbf G}}
\def\bI{{\mathbf I}}
\def\bM{{\mathbf M}}
\def\bY{{\mathbf Y}}
\def\bX{{\mathbf X}}
\def\bS{{\mathbf S}}
\def\bb{{\mathbf b}}
\def\bh{{\mathbf h}}
\def\bg{{\mathbf g}}
\def\bla{{\mathbf \la}}
\def\bmu{\mathbf m }
\def\by{{\mathbf y}}
\def\bmu{\mbox{\boldmath $\mu$} }
\def\bsig{\mbox{\boldmath $\sigma$} }
\def\bunity{{\mathbf 1}}
\def\cA{{\cal A}}
\def\cB{{\cal B}}
\def\cC{{\cal C}}
\def\cD{{\cal D}}
\def\cF{{\cal F}}
\def\cG{{\cal G}}
\def\cH{{\cal H}}
\def\cI{{\cal I}}
\def\cL{{\cal L}}
\def\cN{{\cal N}}
\def\cM{{\cal M}}
\def\cO{{\cal O}}
\def\cR{{\cal R}}
\def\cS{{\cal S}}
\def\cT{{\cal T}}
\def\eV{{\rm eV}}
%




\large
 \begin{center}
 {\Large \bf Un-oriented Quiver Theories for Majorana Neutrons} \
 \end{center}

 \vspace{0.1cm}

 \vspace{0.1cm}
 \begin{center}
{\large Andrea Addazi}
\footnote{E-mail: \,  andrea.addazi@infn.lngs.it}
\end{center}
{\it Dipartimento di Fisica,
 Universit\`a di L'Aquila, 67010 Coppito AQ and
LNGS, Laboratori Nazionali del Gran Sasso, 67010 Assergi AQ, Italy}

 \begin{center}
{\large Massimo Bianchi}
\footnote{E-mail: \, massimo.bianchi@roma2.infn.it}
\end{center}
{\it Dipartimento di Fisica, Universit\'a di Roma ÒTor VergataÓ
I.N.F.N. Sezione di Roma ÒTor VergataÓ, Via della Ricerca Scientifica, 1
00133 Roma, ITALY}

\vspace{1cm}
\begin{abstract}
\large

In the context of un-oriented open string theories, 
we identify quivers whereby 
a Majorana mass for the neutron is indirectly generated by exotic instantons.  
We discuss two classes of (Susy) Standard Model like 
quivers, depending on the embedding of $SU(2)_W$ in the Chan-Paton group.
In both cases, the main mechanism involves a vector-like pair 
mixing through a non-perturbative mass term. 
We also discuss possible relations between the phenomenology of 
Neutron-Antineutron oscillations and LHC physics in these models.
In particular, a vector-like pair of color-triplet scalars or color-triplet fermions 
could be directly detected at LHC, compatibly with $n-\bar{n}$ limits. 
Finally we briefly comment on Pati-Salam extensions of our models.

\end{abstract}

\baselineskip = 20pt

\section{Introduction}

Recently we have proposed the possibility that a Majorana mass term for the neutron could be indirectly generated  
by non-perturbative quantum gravity effects present in string theory: the {\it exotic instantons} \cite{Addazi:2014ila,Addazi:2015ata} \footnote{For the classification of instanton effects in strings theory see: \cite{uno}-\cite{sei}  for world-sheet
instantons in the heterotic string, \cite{sette}-\cite{nove} for E2-instantons in the Type IIA string,
\cite{dieci}-\cite{dodici} for M2-brane and M5-brane instantons in M-theory, \cite{tredici}-\cite{quindici}  for the $D3$-$D(-1)$ system in type IIB, \cite{sedici} for the effect of background fluxes on $E2$-instantons, \cite{diciassette} for $E3$-instantons in Type IIB theory. In \cite{Bianchi:2009bg,Bianchi:2007fx}, instantons in Z3 orbinfolds are discussed. }.
In theories with open and un-oriented strings, instantons have a simple geometrical interpretation: they are nothing but Euclidean D-branes (a.k.a. E-branes) wrapping internal cycles of the compactification. 
`Gauge' instantons are E-branes wrapping the same cycles as some D-branes present in the vacuum configuration. `Exotic' stringy instantons are E-branes wrapping cycles different from those wrapped by the D-branes present in the background. For `gauge' instantons, a very natural and intuitive embedding of the ADHM data \cite{ADHM} is represented by open strings with at least one end on the E-branes. Exotic instantons admit a similar description that however escapes the ADHM construction in that -- in the simplest and most interesting case -- the moduli are purely fermionic. Non-perturbative effects generated by both `gauge' and `exotic' instantons are calculable in string inspired extensions of the (supersymmetric) Standard Model. Exotic instantons can break both anomalous axial symmetries and vectorial ones. Gauge instantons can break anomalous symmetries only. 

Recall that in the SM only $B-L$ is non-anomalous. Baryon and Lepton numbers are separately anomalous and can be broken by non-perturbative finite-temperature instanton-like effects due to  `sphalerons' \cite{Klinkhamer:1984di}. At low temperature, as in the present cosmological epoch, sphaleron effects are highly suppressed but $SU(2)$ EW `sphalerons' play an important role during the early stages of the universe, up to the electro-weak phase transition.
 In the primordial thermal bath $(B-L)$-preserving transitions can be induced
 by sphalerons because of the thermal fluctuations in the weakly coupled plasma.
They have only the net effect to convert $B$ to $L$ and {\it vice versa}: they cannot provide separate mechanisms for Baryogenesis or Leptogenesis 
 without Physics Beyond the Standard Model \cite{sphalerons}. As explicitly noticed by t'Hooft, these
 $(B-L)$-preserving transitions are suppressed by factors of order 
 $10^{-120}$ and are thus absolutely impossible to detect in the laboratory \cite{tHooft}. 
 (See also \cite{Sh} for a classical textbook on these aspects).
 
On the other hand, `exotic' instantons can also break vector-like and non-anomalous symmetries and in principle they can be {\it unsuppressed}. This peculiar feature of exotic instantons can lead to interesting $B$- (or $L$-) 
and $B-L$ violating physics
testable in laboratories, such as a Majorana mass for the neutron and related $n-\bar{n}$ oscillations \cite{Addazi:2014ila}. 
These effects can also dynamically propagate from the string scale
to much lower energies, as shown in \cite{Addazi:2014ila}
\footnote{For other recent developments of this idea see also 
\cite{Addazi:2015eca,Addazi:2015fua,Addazi:2015oba,Addazi:2015goa}.}.
The possibility of an effective Majorana mass term for the neutron was firstly proposed by Majorana himself in \cite{1}. Such a mass term could induce neutron-antineutron transitions, violating Baryon number, contrary to the predictions of the Standard Model \cite{nnbar}. The next generation of experiments is expected to test $PeV$ physics \cite{ProjectX, Phillips:2014fgb} by improving the limits on the oscillation time to $\tau_{\rm n-\bar{n}}\simeq 10^{10}\, \rm s$, 
two orders of magnitude higher than the current limits \cite{Baldo}. 

Let us suppose instead that $n-\bar{n}$ oscillations be found in the next run of experiments: then it would be challenging to generate such an effect with a time-scale
around $10^{10}\,{\rm s} \approx 300\,{\rm yr}$ without fast proton decay ($\tau_{\rm p} > 10^{34}\,\rm yr$) or unsuppressed FCNC's. 
The class of models that we consider seems to meet these requirements.
Exotic instantons propagate the quantum gravity stringy effects to much lower scales,
that can be as low as $1000\, \rm TeV$. 

The main purpose of the present paper is to clarify aspects of  
the mechanism proposed in  \cite{Addazi:2014ila} 
and to identify {\it quiver theories} leading 
to the interesting phenomenology introduced in  \cite{Addazi:2014ila}. 
The paper is organized as follows:
in Section 2 we briefly review the main features of the models with Majorana mass terms for the neutron; in Section 3, we discuss the phenomenology related to neutron-antineutron oscillations, reviewing and extending our previous considerations
and constraining the allowed region in parameter space, with particular attention to possible signatures at LHC; in Section 4 we briefly review the construction of (un-oriented) quiver theories and identify SM-like (un-oriented) quivers for a Majorana neutron; 
in Section 5 we discuss possible quantum corrections to the K\"ahler potential and D-terms 
as well as the role of susy breaking;  
in Section 6 we present our conclusions and a preliminary discussion of Pati-Salam extensions presented in \cite{Addazi:2015hka}. 

\section{A simple class of models}

The models we consider are based on D6-branes wrapping 3-cycles in $CY_{3}$ and
giving rise to such gauge groups as $U(3)\times U(2)\times U(1)_{1} \times U(1)'$ or $U(3)\times Sp(2)\times U(1)_{1} \times U(1)'$. We will also need 
$\Omega$-planes for local tadpole cancellation and $E{2}$-branes (instantons). The un-oriented strings between the various stacks account for the minimal super-field content of the MSSM
\begin{equation}\label{super}
Q_{+1/3}^{i,\alpha}\,\,\, L_{-1}^{\alpha}\,\,\,U^{c}_{i,-4/3}\,\,\, E^{c}_{+2}\,\,\, D^{c}_{i,+2/3}\,\,\,H_{u,+1}^{\alpha}\,\,\,H_{d,-1}^{\alpha}
\end{equation}
where we indicated explicitly the hyper-charge of the various super-fields. These interact via the super-potential 
\begin{equation}\label{WU}
\mathcal{W}=y_{d}H_{d}^{\alpha}Q_{\alpha}^{i}D_{i}^{c}+y_{l}H_{d}^{\alpha}L_{\alpha}E^{c}+y_{u}H_{u}^{\alpha}Q_{\alpha}^{i}U_{i}^{c}+\mu H_{u}^{\alpha}H_{\alpha d}
\end{equation}
Flavour or family indices are understood unless strictly necessary.
Note that $\mathcal{W}$ preserves R-parity. The last term violates the continuous R-symmetry and can be generated by $E{2}$-branes (instantons) as discussed in \cite{Blumenhagen:2006xt,Ibanez1,Ibanez2} or by `supersymmetric'  bulk  fluxes as reviewed later on.

We could also consider some of the possible perturbative R-parity breaking terms (see \cite{Barbier} for a review on the subject): 
\begin{equation}\label{WR}
\mathcal{W}_{RPV}=\lambda_{_{LLE}}L^\alpha L_\alpha E^{c}
+\lambda'_{_{LQD}}L^\alpha Q^i_\alpha D^{c}_i + \lambda''_{_{UDD}} \epsilon^{ijk} U^c_{i}D^{c}_{j}D^{c}_{k}+\mu_{_{LH}}L^\alpha H^u_\alpha
\end{equation}

Moreover, soft susy breaking terms can be generated by  `non-supersymmetric bulk' fluxes or other means, that produce scalar mass terms, Majorana mass terms for gaugini (zino, photino, gluini), trilinear $A$-terms, bilinear  $B$-terms \cite{F1, F2}. 

In the first case,  the hypercharge $Y$ is a combination of the four $U(1)$ charges\begin{equation}\label{gaugegroup}
U(3)\times U(2) \times U(1)\times U(1)' \simeq SU(3)\times SU(2) \times U(1)_{3} \times U(1)_{2} \times U(1)\times U(1)'
\end{equation}
In fact the four $U(1)$'s can be recombined into $U(1)_{Y}$, $U(1)_{B-L}$ and two anomalous $U(1)$'s.  In the other case, with gauge group $U(3)\times Sp(2) \times U(1)\times U(1)'$, 
one has 
\begin{equation}\label{gaugegroup2}
U(3)\times Sp(2) \times U(1)\times U'(1) \simeq SU(3)\times SU(2) \times U(1)_{3} \times U(1)\times U(1)'
\end{equation}
and $Y$ is a linear combination of $q_{1,1',3}$. 

The presence of anomalous $U(1)$'s is not a problem in string theory.  
A generalisation of the Green-Schwarz mechanism disposes 
of anomalies. In particular in the string-inspired extension of the (MS)SM under consideration, new vector bosons $Z'$ appear
that get a mass via a St\"uckelberg mechanism \cite{145} and interact through generalized Chern-
Simon (GCS) terms, in such a way as to cancel all anomalies \cite{146,147,Bianchi:2007fx}
\footnote{We mention that stringent limits on another application of the St\"uckelberg mechanism in massive gravity were discussed 
in \cite{Addazi:2014mga}. }. 

If the relevant D-brane stacks intersect four rather than three times, {\it i.e.} $\#U(3)\cdot U(1)=4$, a {\it 4th} replica $D'=D^{c}_{f=4}$ of the three MSSM $D^{c}_{f=1,2,3}$ appears. 
 Moreover, compatibly with tadpole and anomaly cancellation, another chiral super-field $C^{i}=\frac{1}{2}\epsilon^{ijk}C_{jk}$ appears at the intersection of the two images of the $U(3)$ stack of $D6$-branes, reflecting each other on the $\Omega$-plane\footnote{Note that the first two ingredients -- MSSM super-fields and R-preserving super-potential -- have been widely explored in the literature, 
the additional vector-like pair and the $\Omega$-plane mark the main difference between our model, proposed in \cite{Addazi:2014ila}, and the ones already known.}. 

$D'^{c}_{Y=+2/3}(B=-1/3)$ and $C_{Y=-2/3}^{i}(B=-2/3)=\frac{1}{2}\epsilon^{ijk}C_{jk}$ form a vector-like pair with respect to $SU(3)\times U(1)_Y$. For the moment, we just introduce such a vector-like pair 
of superfields {\it ad hoc}, we will later see how the precise hyper-charge and baryon number assignments arise in the D-brane context. With 
the desired choice of hypercharges,  
such a pair does not introduces anomalies 
for the SM gauge group\footnote{For instance, no extra $SU(3)^{3}$ 
anomalies are introduced since, for $N_c=3$,
$Tr_{N_{c}(N_c-1)/2}(T^aT^bT^c) = (N_{c}-4) d^{abc}= -d^{abc}= - Tr_{N_c}(T^aT^bT^c)$.}. 
In the following, we will show appropriate examples of (un-)oriented quivers, 
that can possibly embed such a model.

Compatibly with gauge invariance,
one can introduce new perturbative Yukawa-like interactions of $D'$ and $C$:
\begin{equation}\label{W1}
\mathcal{W}_{1}=h_{D'}Q^{\alpha i}H^d_{\alpha}D_{i}^{'c}\end{equation} 
and 
\begin{equation}\label{W2}
\mathcal{W}_{2}=h_{C}Q^{i}Q^{j}C_{ij}
\end{equation}
A non-perturbative mixing mass term 
\begin{equation}\label{Wexotic}
\mathcal{W}_{exotic}={1\over 2} \mathcal{M}_{0}\epsilon^{ijk}D_{i}^{'c}C_{jk}\end{equation}
can be generated by non-perturbative $E2$-instanton effects. The relevant E2-brane (exotic instanton) is transversely invariant under $\Omega$ and intersects the {\it physical} D6-branes, as discussed in \cite{Addazi:2014ila}. The non-perturbative mass scale is $\mathcal{M}_{0}\sim M_{S}e^{-S_{E2}}$
with $M_{S}$ the string scale, $S_{E2}$ the $E{2}$ instanton action, depending 
on the closed string moduli parameterizing the complexified size of the 3-cycle wrapped by the world-volume of $E2$.

Integrating out the vector-like pair an effective super-potential of the form
\begin{equation}\label{effective}
\mathcal{W}_{eff}=h_{D'}h_{C} \frac{1}{\mathcal{M}_{0}}Q_{f_{1}}^{\alpha i}H^d_{\alpha}Q_{\beta f_{2}}^{j}Q^{k\beta}_{f_{3}}\epsilon_{ijk}
\end{equation}
is generated.
 
In this way, one can start with a theory preserving R-parity and have it broken dynamically only through the non-renormalizable R-parity breaking operator (\ref{effective}).

In principle, one can also consider some explicit R-parity breaking terms, including perturbative ones (\ref{WR}), but then one has to carefully study the dangerous effect of these on low-energy processes
violating baryon and lepton numbers. 

\begin{table}[htbp]
\centering 
\begin{tabular}{||l|rl|l||}
\hline
\parbox{0.2\textwidth}{Super-fields}  &  \parbox{0.1\textwidth}{$\,\,\,\,B$} & \parbox{0.1\textwidth}{$\,\,\,\,L$}      \\
\hline
\parbox{0.2\textwidth}{$D'_c(\bar{3},1;+2/3)$}   
  &    \parbox{0.1\textwidth}{$-1/3$} &   \parbox{0.1\textwidth}{$\,\,\,\,0$}  \\
\parbox{0.2\textwidth}{$C(3,1;-2/3)$}  &    \parbox{0.1\textwidth}{$-2/3$} & \parbox{0.1\textwidth}{$\,\,\,\,0$}      \\
\parbox{0.2\textwidth}{$H_{u}(1,2;+1)$}   & \parbox{0.1\textwidth}{$\,\,\,\,\,\,0$} & \parbox{0.1\textwidth}{$\,\,\,\,0$} \\
\parbox{0.2\textwidth}{$H_{d}(1,2;-1)$}   & \parbox{0.1\textwidth}{$\,\,\,\,\,\,0$} & \parbox{0.1\textwidth}{$\,\,\,\,0$} \\
\parbox{0.2\textwidth}{$Q(3,2;+1/3)$}    & \parbox{0.1\textwidth}{$+1/3$} & \parbox{0.1\textwidth}{$\,\,\,\,0$} \\
\parbox{0.2\textwidth}{$U^c(\bar{3},1;-4/3)$}    & \parbox{0.1\textwidth}{$-1/3$} & \parbox{0.1\textwidth}{$\,\,\,\,0$} \\
\parbox{0.2\textwidth}{$D^c(\bar{3},1;+2/3)$}    & \parbox{0.1\textwidth}{$-1/3$} & \parbox{0.1\textwidth}{$\,\,\,\,0$} \\
\parbox{0.2\textwidth}{$L(1,2;-1)$}   & \parbox{0.1\textwidth}{$\,\,\,\,\,\,0$} & \parbox{0.1\textwidth}{$+1$} \\
\parbox{0.2\textwidth}{$E^c(1,1;2)$}   & \parbox{0.1\textwidth}{$\,\,\,\,\,\,0$} & \parbox{0.1\textwidth}{$-1$} \\
\hline
\end{tabular}
\caption{ In the table, we summarise the matter super-field content in our model.
In particular, we introduce one extra vector-like pair $D_c',C$ with respect to the SM
super-fields. 
We list their representations with respect to SM gauge group
$SU(3)\times SU(2) \times U(1)_{Y}$, 
their Baryon and Lepton numbers $B,L$.   }
\end{table}

\section{Phenomenology: Neutron-Antineutron physics and LHC}

\begin{figure}[t]
\centerline{ \includegraphics [height=6cm,width=0.5 \columnwidth]{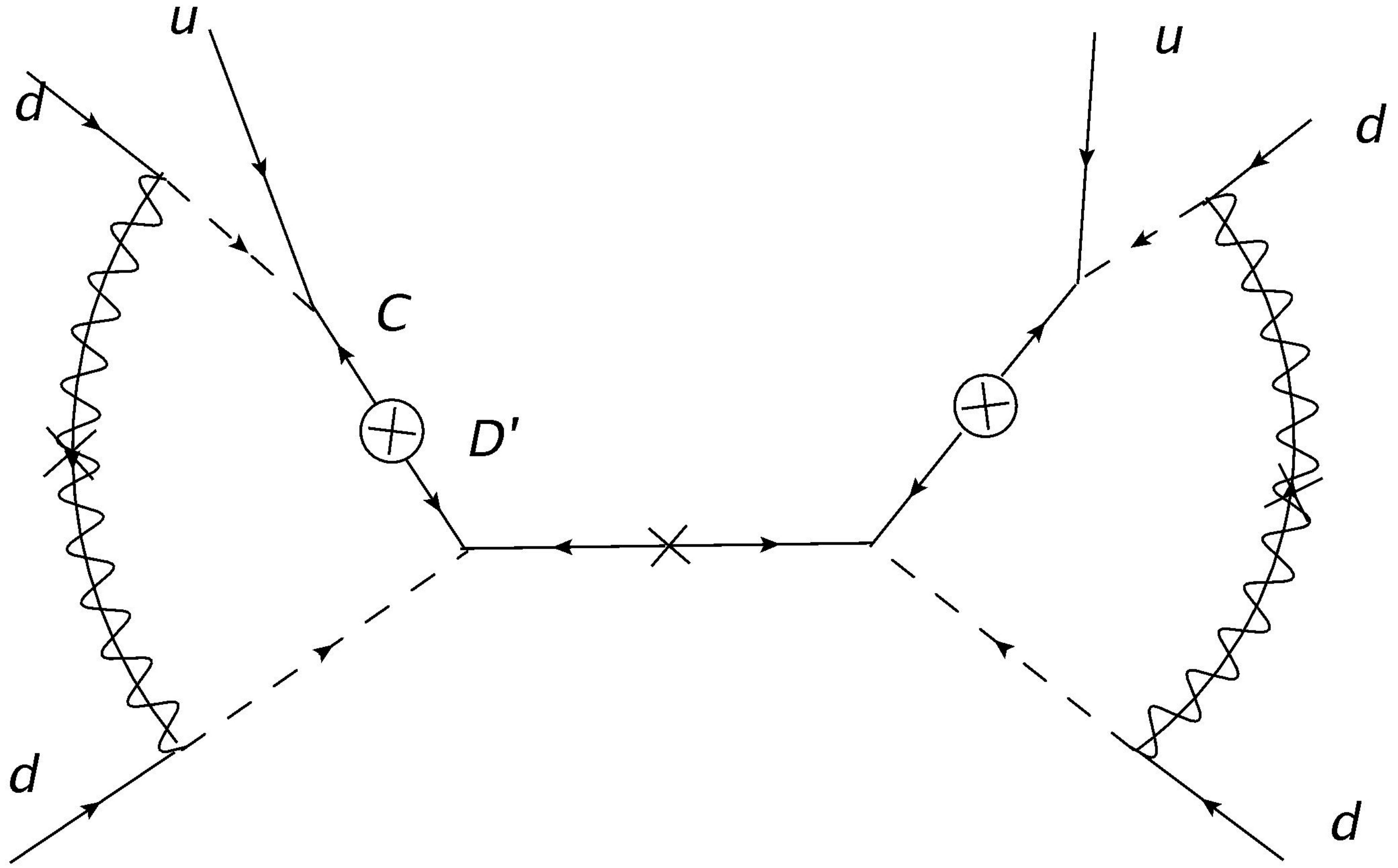}}
\vspace*{-1ex}
\caption{Diagram inducing neutron-antineutron transitions through  vector-like pair
  of fermions $D',C$ (the white blob represent the non-perturbative mass term induced by exotic instantons), an Higgsino, and a conversion of squarks into quarks through gaugini, like zini or gluini. }
\label{plot}   
\end{figure}

\begin{figure}[t]
\centerline{ \includegraphics [height=5cm,width=0.6\columnwidth]{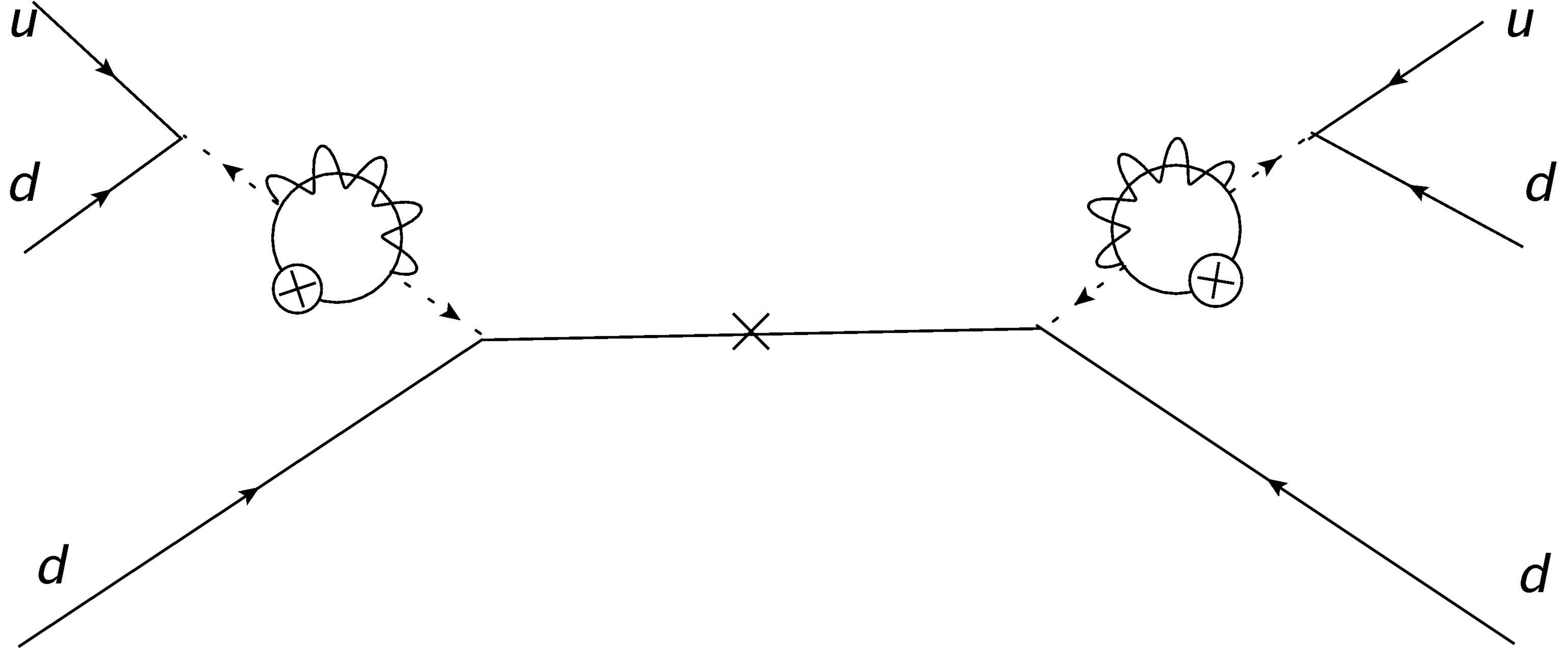}}
\vspace*{-1ex}
\caption{Diagram inducing neutron-antineutron transitions through vector-like pair
  of scalars and an Higgsino. The scalars of the superfields $D'-C$ are mixing through a loop 
  of their fermionic superpartners (the white blob represent the non-perturbative mass term induced by exotic instantons) and a gaugino. }
\label{plot}   
\end{figure}

An operator like (\ref{effective}) generates neutron-antineutron transitions, violating baryon number with $\Delta B=2$, as shown in Fig.1 and discussed in \cite{Addazi:2014ila}. 
The scale $\mathcal{M}_{n\bar{n}}^{5}=m_{\tilde{g}}^{2}\mathcal{M}_{0}^{2}M_{\tilde{H}}$ in $(udd)^{2}/\mathcal{M}_{n\bar{n}}^{5}$
is a combination of the gaugino (gluino or zino) mass $m_{\tilde{g}}$,
of the mixing mass term $\mathcal{M}_{0}$ for the vector-like pair  and the Higgsino mass $M_{\tilde{H}}$.
In order to satisfy the present experimental bound $\mathcal{M}_{n\bar{n}}>300\, \rm TeV$, we can consider different scenarios. We focus on some of these in the following\footnote{ In \cite{Addazi:2014ila}, we have made the tacit and not fully justified assumption that 
the gaugino mass were $m_{\tilde{g}}\simeq \mathcal{M}_{0}$. Here, we relax this assumption. }.
i) Higgsini, Gaugini and vector-like pairs at the same mass scale $300-1000\, \rm TeV$,
in order to trivially satisfy the bound;
ii) Susy breaking at the TeV scale, with $M_{\tilde{H}}\simeq M_{\tilde{g}}\simeq 1\, \rm TeV$, and $\mathcal{M}_{0}\sim 10^{15\div 16}\, \rm GeV$;
iii) Heavy Higgsini and gaugini: $M_{\tilde{g}}\simeq M_{\tilde{H}}\simeq M_{SUSY}\simeq 10^{4}\, \rm TeV$, 
$\mathcal{M}_{0}\simeq 1\, \rm TeV$ \footnote{In \cite{Addazi:2015dxa}, a string-inspired non-local susy QFT model 
was studied, while in \cite{Addazi:2015ppa} the formation of non-perturbative classical configurations in scattering amplitudes of an effective non-local QFT 
was studied, with possible connections with exotic instantons' productions in collisions. }. 

Another diagram generating $n-\bar{n}$ transitions is depicted in Fig.2, the analysis of the parameter space is 
roughly the same as for the first case. 

These diagrams respect R-parity at all the vertices, except for the 
non-perturbative mixing term of the vector-like pair.
In fact the super-potential has $R(\mathcal{W})=-1$ as usual,  
and one can consistently assign R-charges to $C$ and $D'$, 
so that their tri-linear Yukawa terms be invariant. Yet their mass term necessarily violates R-parity. Omitting the coupling constants one schematically has
\begin{equation}\label{Rcharge}
\mathcal{L}_y^{-}=\psi_{C}^{-}\tilde{q}^{-}q^{+}+\phi_{D'}^{-}q^{+}\psi_{H_{d}}^{-} + \phi_{C}^{+}q^{+}q^{+}+ \psi_{D'}^{+}q^{+}\phi_{H_{d}}^{+} + \mathcal{M}_{0}\psi_{C}^{-}\psi_{D}^{+}
\end{equation}
where $\pm$ indicates the R-parity, $\phi_{C,D'}$ and $\psi_{C,D'}$ are the 
scalars and the fermions respectively in the superfields $C,D'$, 
$q,\tilde{q}$ are quarks and squarks, $\phi_{H_{u,d}}$ are Higgs bosons, 
$\psi_{H_{u,d}}$ are the two Higgsini. Note how R-parity is violated only by the last non-perturbative term with mixing mass parameter $\mathcal{M}_{0}$ not directly connected to 
the Dirac mass term for $D'$, emerging from its `standard' coupling to the Higgs. 

More precisely, $\mathcal{M}_{0}$ is replaced by the mass parameter of the lightest mass                                                 state, be it a  
fermion $\psi_{D',C}$ as in Fig.1, or a scalar $\phi_{D',C}$ as in Fig.2. 
The scalars $\phi_{D',C}$ have in general 
a non-diagonal mass matrix \cite{Addazi:2015ata} of the form 
\begin{equation}\label{MassMat}
 M^{2}_{\rm b} = \left( \begin{array}{cccc} m_{\phi_{D'}}^{2} & 0 & \delta \mu^{2} & 0
\ \\ 0 & m_{\phi_{D'}}^{2} & 0 & -\delta \mu^{2} \ \\
\delta \mu^{2} & 0 & m_{\phi_{C}}^{2}  & 0 \ \\
0 & -\delta \mu^{2} & 0 & m_{\phi_{C}}^{2}
\end{array} \right)
\end{equation} 
written in the basis $(\phi^{1}_{D'},\phi^{2}_{D'},\phi^{1}_{C},\phi^{2}_{C})$, 
with $\phi_{D',C}=\phi^{1}_{D',C}+i\phi^{2}_{D',C}$ (assuming $\delta \mu=\delta \mu^{*}$),
and 
with
\begin{equation}\label{scal}
\mathcal{L}_{m}=m_{\phi_{D'}}^{2}\phi_{D'}^{\dagger}\phi_{D'}+m_{\phi_{C}}^{2}\phi_{C}^{\dagger}\phi_{C}+h.c
\end{equation}
and $\delta \mu^2 \sim m_{\tilde{g}}\mathcal{M}_{0}$ as in Fig.2. 

The mass eigenvalues of (\ref{MassMat}) are
\begin{equation}\label{e12}
 \lambda_{\pm}^{2}=\frac{1}{2}\left(m_{\phi_{D'}}^{2}+m_{\phi_{C}}^{2} \pm \sqrt{4\delta \mu^{4}+(m_{\phi_{D'}}^{2}-m_{\phi_{C}}^{2})^{2}}\right)
\end{equation}
both doubly degenerate, as manifest in (\ref{MassMat}). Note that, in the case of $m_{\phi_{D'}}=m_{\phi_{C}}=0$
and $\delta \mu \neq 0$, 
one of the mass eigenvalues is negative, {\it i.e.} leading to a condensate, breaking $SU(3)_{c}$. 

On the other hand, we would like to note that Dirac mass terms for fermions $\psi_{D'}$ and $\psi_{C}$ 
are not present at all. For instance, $\psi_{D'}$ is like a 4th right-handed down quark without a Left-Handed counterpart. As a result $m_{\pm}=\pm \mathcal{M}_{0}$, where the sign, in fact any phase, can be absorbed into a redefinition of the phases of the fermionic fields.

We can distinguish two branches for LHC and FCNC's phenomenology:
i) First Susy hierarchy; ii) Second Susy hierarchy.

In the {\it First Susy hierarchy}, scalars $\phi_{D',C}$ are the lighest mass eigenstates
 $\lambda_{-}<<|m_{-}|$, {\it i.e.} scalars have lower masses with respect to their fermionic partners.
 This case corresponds to an ordinary susy hierarchy for $C$ and to an inverted  
 susy hierarchy for $D'$.  In this case, the relevant contribution for Neutron-Antineutron oscillations is the one in Fig.2.
In string inspired models scalars can get extra contributions from non-supersymmetric closed-string fluxes (NS-NS or R-R),
not contributing to fermion masses\footnote{See for example \cite{F1,F2}, for discussions about
soft susy breaking terms generated by fluxes in MSSM's in (unoriented) open string theories.}. Thus inverting the hierarchy for $D'$, {\it i.e.} getting $m_{\phi_{D'}} < m_{\psi_{D'}}$, may required the inclusion of loop effects.
In principle, $\mathcal{M}_{0}m_{\tilde{g}}$ has to be replaced by the lighest mass $\lambda_{-}^{2}$, in the parametric estimate for $n-\bar{n}$ oscillations shown above.
For $m_{\phi_{C}}^{2}>>m_{\phi_{D'}}^{2}\simeq m_{\tilde g}\mathcal{M}_{0}$, we obtain $\lambda_{-}\simeq \mathcal{M}_{0}$, and dangerous FCNC's can be suppressed if $m_{\phi_{C}}>>m_{\phi_{D'}}$  \cite{Addazi:2015ata}. 
In particular, assuming $m_{\phi_{C}}^{2}\simeq 10^{6}m_{\phi_{D'}}^{2}$
and $\mathcal{M}_{0}^{2}\simeq m_{\phi_{D'}}^{2}$, we obtain, from (\ref{e12}): 
$\lambda_{-}^{2}\simeq m_{\phi_{D'}}^{2}$ and $\lambda_{+}^{2}\simeq m_{\phi_{C}}^{2}$,
with mixing angles $\theta_{13}=\theta_{24}\sim 10^{-6}$.
So, mixings between $\phi_{C}$ and $\phi_{D'}$ are strongly suppressed 
in this case, but may be enough for neutron-antineutron transitions:
a prefactor of $10^{-12}$ in a $n-\bar{n}$ scale $(\mathcal{M}_{0}^{4}\mu)^{1/5}$
has to be included.
This drastically changes the constraints on the other parameters:
for $\mathcal{M}_{0}=1-10\, \rm TeV$, a light $\psi$ of $\mu=1\div 100\, \rm GeV$ would be enough!
The phenomenology of (i) is discussed in \cite{Addazi:2015ata}.
In \cite{Addazi:2015ata}, a toy-model was shown 
in which the so called $\mathcal{X},\mathcal{Y}$ are nothing
but $\phi_{D',C}$ respectively. There are some differences not allowing a perfect identification 
$\mathcal{X}=\phi_{D'}$ and $\mathcal{Y}=\phi_{C}$.
For example, the main interactions terms are $\mathcal{Y}u^{R}d^{R}$ rather than $\phi_{C}u_{L}u_{R}$,
and $\mathcal{X}d_{R}\psi$ rather than $\phi_{C}d_{L}\tilde{H}$,
with $\psi$ a sterile Majorana fermion with zero hyper-charge rather than an Higgsino.
This leads to different hyper-charge assignments (reversed in the vector-like pair) and Baryonic number assignment (opposite sign in both), compatibly with gauge symmetries. 
Even so, the phenomenology is very similar to the one discussed in  \cite{Addazi:2015ata}, in so far as neutron-antineutron
oscillations, LHC signatures, FNCN's, and Post-Sphaleron Baryogenesis are concerned. 
For instance, the lightest mass eigenstate scalars can have $\lambda_{-}\simeq 1\, \rm TeV$,
with possible detection channels at LHC. In particular, $pp\rightarrow jjE_{T}\kern-12pt\slash\,\,\,\,$ 
is expected, avoiding stronger constraints from FCNC's processes.
Our models also predict $pp\rightarrow 4j$ (direct bound of $1.2\, \rm TeV$) or $pp\rightarrow t\bar{t}jj$ (direct bound of $900\, \rm GeV$), but FCNC bounds are stronger than LHC ones, in these cases. 

 In the {\it Second Susy Hierarchy}, we consider the opposite scenario in which $\mathcal{M}_{0}<<\lambda_{-}$,
{\it i.e.} susy fermions $\psi_{D',C}$ are lighter than scalars $\phi_{D',C}$. 
In this case, $D'$ has a normal susy hierarchy, while $C$ has an inverted one. 
An inverted hierarchy can be understood as an effect of non-supersymmetric fluxes that give extra contributions to scalar masses with respect to their susy partners.
In this case, direct detection of 
$\psi_{C}-\psi_{D'}$ at LHC would be possible. Their production is possible in several different processes, having peculiar decay channels like $\psi_{C} \rightarrow q\tilde{q}$. 
We would like to note that in our case one can also generate perturbative Yukawa coupling of $\psi_{D'}$ with the bottom quark, leading to a decay channel $\psi_{D'} \rightarrow Hb$. 
Moreover, an electroweak mixing with the top quark is also possible, that would
lead to $\psi_{D'}\rightarrow Wt$. These could be interesting for LHC. 
The present limits on these rare processes are shown in ATLAS EXPERIMENT Public Results in the section devoted to Exotics \cite{Exotic}. The limits on the mass of an additional vector-like pair are of order $500-850\, \rm GeV$. 

In summary, thanks to the non-perturbative R-parity breaking mixing mass term $\mathcal{M}_{0} \epsilon^{ijk}C_{ij}D'_{k}$, the phenomenology of our model is somewhat different with respect to the one of other models with vector-like pairs.
On the other end, a low mass higgsino and consequently a low mass LSP neutralino, detectable at LHC,
would be possible if the mass of the vector-like pair $\mathcal{M}_{0}$ were around $10^{12 \div 15}\, \rm TeV$. This last scenario can be compatible with susy breaking scale around some TeV's.

\subsection{Dangerous operators: no proton decay without right-handed Majorana neutrini}

Let us consider a complete and extended super-potential consisting of the R-parity preserving Yukawa terms of the MSSM (\ref{WU}), the new perturbative Yukawa terms of $C$ and $D$, the non-perturbative mixing mass term of the vector-like pair, and the interaction terms of an extra Right-Handed neutrino $N$, with Majorana mass term and perturbative Yukawa term:
\begin{equation}\label{W}
\mathcal{W}=y_{u}H_{u}QU^{c}+y_{d}H_{d}QD^{c}+y_{l}H_{d}LE^{c}+y_{N}H_{u}LN^{c}+\mu H_{u}H_{d}
\end{equation}
$$+\frac{1}{2}m_{N}N^{2}+h_{C}CQQ+h_{D'}H_{d}QD'^{c}+\mathcal{M}_{0}D'C$$
It is technically natural to neglect other R-parity violating terms, allowed by gauge symmetry. Indeed they do not arise in the quivers discussed later.

In order to integrate out the massive super-fields, $N$, $H_u$, $H_d$, $C$, $D'$, we have to evaluate the field-dependent mass matrix $M_{IJ}(\Phi)$, where $\Phi $ collectively denotes the light super-fields, and invert it 
\begin{equation} \label{Wefff}
\mathcal{W}_{eff} (\Phi) = {1\over 2} F^I (M_{F})^{-1}_{IJ}(\Phi) F^J
\end{equation}
where $F_I = \{F_{N},F_{H_{u}},F_{H_{d}}, F_{C},F_{D'}\}$ indicate the `massive' F-terms. 
The relevant mass matrix is the inverse of\footnote{For simplicity, couplings and flavour structure are understood since they are not relevant in the subsequent discussion.}
\begin{equation}\label{Matrix}
M_{F} = \left( \begin{array}{ccccc} m_{N} & L & 0 & 0 & 0
\ \\  L & 0 & \mu & 0 & 0  \ \\
0 & \mu & 0 & 0 & Q \ \\
0 & 0 & 0 & 0 & \mathcal{M}_{0} \\\
0 & 0 & Q & \mathcal{M}_{0} & 0 
\end{array} \right)
\end{equation}
Due to the non-trivial dependence on the superfields $Q$ and $L$, 
direct inversion of (\ref{Matrix}) becomes laborious but straight-forward 
with the result 
\begin{equation}\label{Matrix2}
(M_{F})^{-1} = \left( \begin{array}{ccccc} \frac{1}{m_{N}} & 0 & -\frac{L}{m_{N}\mu} & \frac{LQ}{m_{N}\mathcal{M}_{0}\mu} & 0\ \\
0 & 0 & \frac{1}{\mu} & -\frac{Q}{\mathcal{M}_{0}\mu} & 0 \ \\
-\frac{L}{m_{N}\mu} & \frac{1}{\mu} & \frac{L^{2}}{m_{N}\mu^{2}} & -\frac{L^{2}Q}{m_{N}\mathcal{M}_{0}\mu^{2}} & 0 \ \\
\frac{LQ}{m_{N}\mathcal{M}_{0}\mu} & -\frac{Q}{\mathcal{M}_{0}\mu} & -\frac{L^{2}Q}{m \mathcal{M}_{0}\mu^{2}} & \frac{L^{2}Q^{2}}{m \mathcal{M}_{0}^{2}\mu^{2}}& \frac{1}{\mathcal{M}_{0}} \ \\
0 & 0 & 0 & \frac{1}{\mathcal{M}_{0}} & 0\ \\
\end{array} \right)
\end{equation}
A perturbative approach, alternative but equivalent to the exact inversion (\ref{Matrix2})
is reported in Appendix. 

On-shell the F-terms yield 
\begin{equation}\label{FN}
F_{N}=0
\end{equation}
\begin{equation}\label{FHu}
F_{H_{u}}=QU^{c}
\end{equation}
\begin{equation}\label{FHd}
F=QD^{c}+LE^{c}
\end{equation}
\begin{equation}\label{C}
F_{C}=QQ
\end{equation}
\begin{equation}\label{Dp}
F_{D'}=0
\end{equation}
Replacing their expressions into $\mathcal{W}_{eff} (\Phi)$, we obtain the following extra and potentially dangerous operators (relevant coupling constants are omitted for simplicity):
\begin{equation}\label{Ffinally1}
\mathcal{W}_{1st}+\mathcal{W}_{2nd}=\frac{1}{\mu \mathcal{M}_{0}}QQQQU^{c}+\frac{L^{2}}{\mu^{2} m_{N}}(QD^{c}+LE^{c})^{2}
\end{equation}
\begin{equation}\label{Finally3}
\mathcal{W}_{3rd}=\frac{L^{2}Q}{m_{N}\mu^{2}\mathcal{M}_{0}}(QQ)(QU^{c}+LE^{c})
\end{equation}
We report also the only one remaining at the 4th order:
\begin{equation}\label{Ffinally4}
\mathcal{W}_{4th}=\epsilon_{ijk}\epsilon_{i'j'k'}(Q^{i}Q^{j})\frac{Q^{k}L L Q^{i'}}{\mu^{2}m_{N}\mathcal{M}_{0}^{2}}(Q^{j'}Q^{k'})
\end{equation}
In the limit of $m_{N}\rightarrow \infty$, all the dangerous operators
are automatically suppressed. In fact, only $QQQQU^{c}/\mu\mathcal{M}_{0}$ remains, 
but this cannot lead to proton decay, as discussed in \cite{Addazi:2014ila}. 
Also combining such operator with other perturbative ones, one can check that all the resulting effective operators are innocuous: there is no operator leading to a
final state without at least one susy partner (so, no available phase space for proton decay), without violation of any fundamental symmetry 
like charge, spin or fermion number. 

In fact our models may be chosen not to violate Lepton number, by setting $m_N =0$, by turning on fluxes or other means that prevent any $E2$-brane instanton that may generate $m_{N}$ \cite{Billo:2008sp}.
The price to pay is that a type I  see-saw mechanism for the neutrino is not allowed:
we cannot generate a Majorana mass without fast proton decay. 
So, such processes as neutrino-less double-$\beta$ decay would provide evidence against these class of models. Of course, a Dirac mass for the neutrino ${\cal W} = H^\alpha_u L_\alpha N \rightarrow {\cal L}_y = \phi_{H_u}^\alpha \ell^t_\alpha \nu_0$ is allowed if R-H sterile neutrini are present. 

\subsection{Flavour changing neutral currents}
Extra contributions to FCNC's may appear in our models,
mediated by $\phi_{C}$, in normal
susy hierarchy, as cited above. 
But these can be sufficiently suppressed,
compatibly with $n-\bar{n}$ limits. 

Other possible contributions, directly connected to $n-\bar{n}$ transitions, 
are strongly suppressed in our model, 
as discussed in \cite{Addazi:2014ila} (See  Fig. 11-12 in  \cite{Addazi:2014ila}: note that extra quarks-squarks conversions are understood that would further suppress the diagrams by the mass of the gaugini). 
In particular, extra contributions to neutral meson-antimeson oscillations like  $K^{0}-\bar{K}^{0}$ are strongly suppressed, approximately  by a power $\mathcal{M}_{0}^{-4}M_{\tilde{H}}^{-2}M_{SUSY}^{-2}$. The choice of $M_{SUSY}$, {\it i.e.} whether gaugini or squarks give extra suppressions, depends on the diagram under consideration. Also in meson decays into two mesons, 
the suppressions are of the same order: $\mathcal{M}_{0}^{-4}M_{\tilde{H}}^{-4}M_{SUSY}^{-2}$. 

\section{Standard Model like quivers generating a Majorana Neutron}

Our aim, in this section, is to identify possible (un)oriented quiver field theories
for the models introduced above, thus generating a neutron Majorana mass. As discussed above 
the ingredients are un-oriented strings stretched between  
stacks of D6-branes, wrapping 3-cycles in a Calabi-Yau 3-fold $CY_{3}$. Thanks to the local CY condition, the resulting theory preserves $N=1$ supersymmetry. We will also need $\Omega$-planes for anomaly cancellation and $E2$-branes,  wrapping (different) 3-cycles in order to generate some non-perturbative interactions. 

\subsection{What is a quiver field theory?}

In general, a quiver,  a collection of arrows, represents 
a gauge theory, with its matter (super)field  content. 
In a quiver, gauge groups are represented by nodes, 
and the fields are represented as (oriented) lines between the nodes.
Adjoint representations start and end on the same node, bi-fundamental representations $({\bf N}, {\bf \bar{M}})$ or $({\bf \bar{N}}, {\bf {M}})$ connect two different nodes. A common example is
$U(N)\times U(M)$ in the oriented case. In the un-oriented case $SO(N)$ and $Sp(2N)$ arise from nodes invariant under a mirror-like involution $\Omega$, associated to the presence of $\Omega$-planes. In this case, (anti)symmetric representations ${\bf N(N\pm1)/2})$ or ${\bf \bar{N}(\bar{N}\pm1)/2})$ as well as $({\bf N}, {\bf{M}})$ or $({\bf \bar{N}}, {\bf \bar{M}})$ correspond to strings going through the mirror. 
In the non-supersymmetric case, a quiver distinguishes scalars and fermions as different kinds of arrows between nodes. 
On the other hand, in a supersymmetric case, a quiver becomes more economic: arrows are superfields, representing both scalars and fermions, and nodes include gaugini as well as gauge fields\footnote{For extended SUSY models with $\mathcal{N} = 2, 4$ one can either use an $\mathcal{N} = 1$ notation,  with arrows representing chiral multiplets and nodes representing vector multiplets, or an $\mathcal{N} = 2$ notation, with unoriented lines representing hyper-multiplets and nodes representing vector multiplets.} The number of arrows on a line correspond 
to the number of generations or replicas of the same (super)field.
A quiver encodes also the possible interactions: 
closed paths (triangles, quadrangles, etc) that respect the orientation of the arrows, represent possible gauge-invariant super-potential or interaction terms.
An effective low energy description of the dynamics of D3-branes 
at Calabi-Yau singularities can be represented as a quiver field theory. 
In this case, standard D-brane stacks are nodes, lines connecting the nodes are (un)oriented open strings stretching between two D-brane stacks,   Euclidean D-branes (instantons) are represented by extra or unoccupied nodes, `dashed' oriented lines connecting these with the original nodes represent modulini. In the quiver notation, interactions between modulini and standard fields 
also correspond to closed polygons (usually triangles) of lines and dashed lines. 

A large class of $\mathcal{N}=1$ superconformal QFTs can arise form D3-branes transverse 
to CY singularities. Near the horizon, the geometry is $AdS_{5}\times \mathcal{X}$ , where $\mathcal{X}$ 
is an Einstein-Sasaki space, base of a Calabi-Yau cone  \cite{KW, MP, BFHMS, FHMSV}. Quivers can be complicated, by the inclusion of $\Omega$-planes
and flavor branes. These generically break superconformal invariance \cite{FHKPU}. 
Including such elements seems necessary for realistic particle physics model building 
\cite{FHKPU, ABPSS, AIQU, CSU, BKLS}, in open string theories \cite{Sagnotti0,Sagnotti8}. 
The brane system is located at the fixed point of the orientifold involution, and the 
low energy dynamics is described, locally, by an un-oriented quiver theory,
with local tadpole cancellation for its consistency at the quantum level, {\it ~i.e.} absence of chiral gauge and gravitational anomalies.
An interesting example, studied in \cite{BIMP}, is the case 
of 
$C^{3}/Z_{n}$ singularities, in the presence of non-compact D7-branes, fractional D3-branes, 
and $\Omega$-planes. 

\subsection{Explicit examples}

\begin{figure}[t]
\centerline{ \includegraphics [height=8.5cm,width=0.6 \columnwidth]{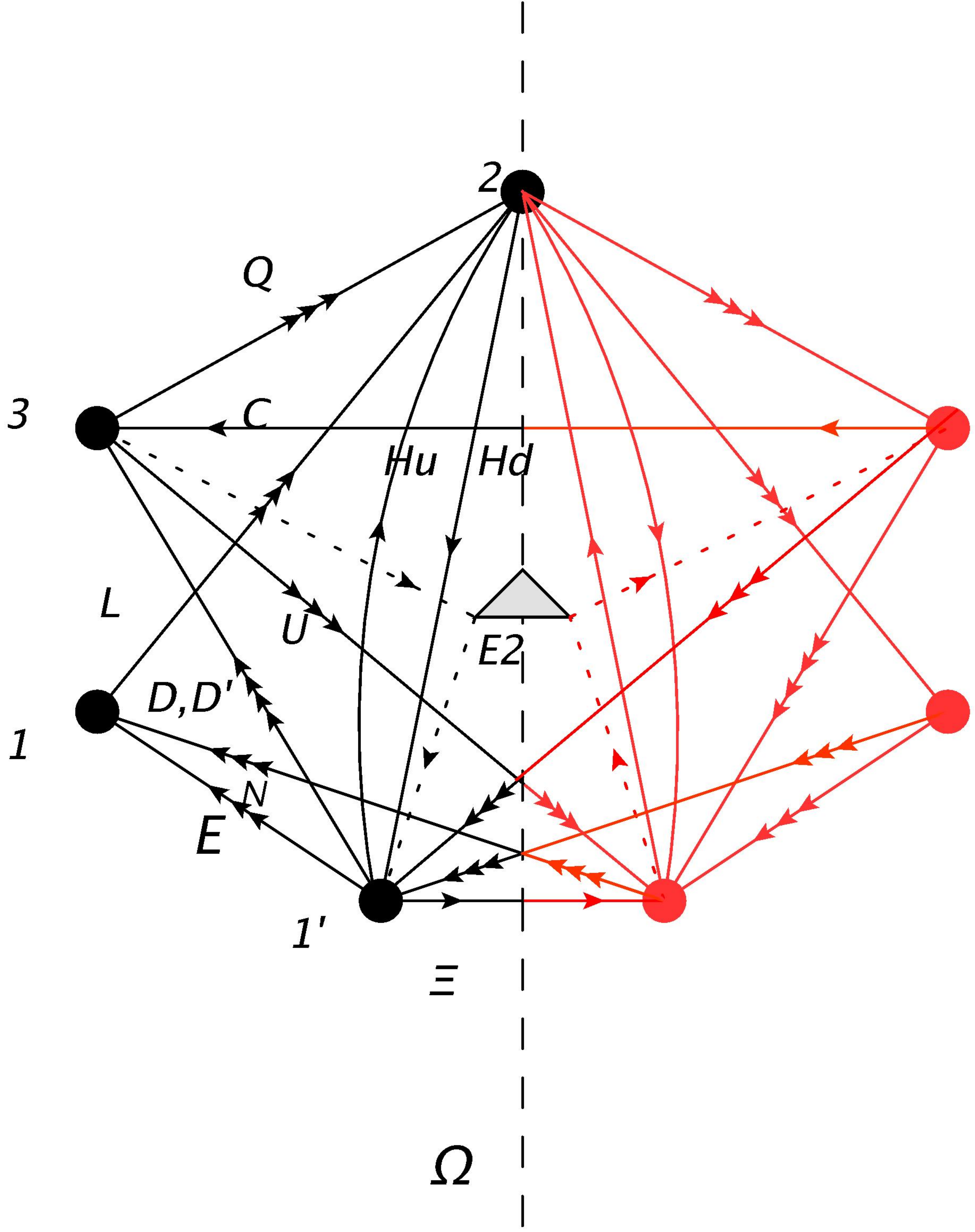}}
\vspace*{-1ex}
\caption{(Susy) Standard Model quiver generating a Majorana mass for the neutron.
}
\label{plot}   
\end{figure}

An example of a simple quiver theory generating a Majorana mass for the neutron is
shown in Fig.~3. This consists in:
one stack of three D6-branes producing the $U(3)$ gauge group, that includes the $SU(3)$ color group and an extra $U(1)$;
two  stacks of single D6-branes producing two $U(1)$ gauge groups, 
an $\Omega$-plane identifying the D-brane stacks 
with their images;  
one stack of two D6-branes, on the $\Omega$-plane,
producing an $Sp(2)$ gauge group, equivalent to the $SU(2)_{L}$ weak group.
As usual in quiver convention, the gauge groups (D-branes stacks) are identified 
with black circles (with a label $3$ for $U(3)$, a label $2$ for $Sp(2)$ and 
labels $1,1'$ for $U(1),U(1)'$); 
the fields, living in the bi-fundamental representation of two gauge groups (strings stretched 
between two stacks of D-branes) are represented as arrows linking the 
two groups involved. 

All the Standard Model super-fields are recovered in 
the present construction. Another check for consistency is to verify that the 
standard Yukawa super-potential terms are 
recovered. In this notation, these corresponds to closed 
circuits (oriented triangles) with sides the super-fields coupled 
via the Yukawa terms. 
For example, it is straightforward to verify that 
$Q, D, H_{d}$ form an oriented triangle 
respecting the orientation of the arrows. 
 So, we recover the standard Yukawa terms
$H_{d}^{\alpha}Q_{\alpha}^{i}D_{i}^{c}$,
$H_{d}^{\alpha}L_{\alpha}E^{c}$,
$H_{u}^{\alpha}Q_{\alpha}^{i}U_{i}^{c}$ with their flavor structure.
The insertion of the {\it 4th} generation 
of D-quarks involves another arrow 
for the consistency of the quiver. 
This arrow is compensated exactly by $C$,
coming from the line between the two images of $U(3)$'s. 
The balance of the arrows is fundamental to have an anomaly-free model
and tadpole cancellation\footnote{It may look suspicious that $\Omega$ acts symmetrical on the D-brane stack producing $Sp(2)$ and anti-symmetrical on the two images of $U(3)$. In the end this is the only choice compatible with the absence of irreducible anomalies. A symmetric irrep (${\bf 6}$) of $SU(3)$ would lead to an inconsistency.}. 
On the other hand, also the new perturbative Yukawa terms
necessary for neutron-antineutron transition 
are generated in our model.
$Q_{i}$ in the left side and $\widehat{Q}_{j}$ in the right side of Fig.3
are closing a triangle with the new exotic field $C^{ij}$ living at the intersection
between $U(3)$ and $\hat{U}(3)$ (with the hat, we 
denote the images in the right side of the $\Omega$-plane in Fig.~3). As a consequence, a perturbative Yukawa term 
$C^{ij}Q_{i}\widehat{Q}_{j}$
is generated. 
On the other hand, $D'^{c}QH_{d}$ is generated exactly 
as the corresponding standard one $D^{c}QH_{d}$. 

Finally, the relevant  exotic $O(1)$ instanton $E2$,
generating the non-perturbative mixing between $D'$ and $C$, is also represented in Fig.~3.
As dashed lines we also denote the modulini living at the intersections 
between $E{2}$ and the $U(3)$ and $U(1)'$ stacks of $D6$-branes. 

The hypercharge in this model is the combination of $3$ charges, 
coming from $U(1)_{3}$, $U(1)$ and $U'(1)$:
\begin{equation}\label{Yper}
Y(Q)=c_3q_{3}+c_1q_{1}+c'_1q'_{1}
\end{equation}
We can fix the coefficients in such a way as to 
recover the standard hypercharges:
\begin{equation}\label{con1}
Y(Q)=\frac{1}{3}=c_3
\end{equation}
\begin{equation}\label{cond2}
Y(U^{c})=-\frac{4}{3}=-c_3-c'_1
\end{equation}
\begin{equation}\label{cond3}
Y(D^{c})=Y(D'^{c})=\frac{2}{3}=-c_3+c'_1
\end{equation}
\begin{equation}\label{cond4}
Y(L)=-1=c_1
\end{equation}
\begin{equation}\label{cond5}
Y(H_{d})=-Y(H_{u})=-1=-c'_1
\end{equation}
\begin{equation}\label{cond6}
Y(E^{c})=2=-c_1+c'_1
\end{equation}
\begin{equation}\label{cond6}
Y(N_{R})=0=-c_1-c'_1
\end{equation}
\begin{equation}\label{cond7}
Y(C)=-\frac{2}{3}=-2c_3
\end{equation}
leading to the result
\begin{equation}\label{result1}
c_3=\frac{1}{3},\,\,\,c_1=-1,\,\,\,c'_1=1 \quad \rightarrow \quad  Y=\frac{1}{3}q_{3}-q_{1}+q_{1'}
\end{equation}

For the quiver in Fig.~3, it is possible to generate a mixing mass term for the Higgses like
$\mu H_{u}H_{d}$ through supersymmetric bulk fluxes since $H_{u}$, $H_{d}$ form a vector-like pair.

The quiver in Fig.~3 is free of tadpoles and the hypercharge $U(1)_{Y}$ is massless. 
As discussed in  \cite{bound1,bound2}, 
a generic quivers has to satisfy two conditions 
in order to be anomalies'/tadpoles' free and 
in order to have a massless hypercharge.
The first one associated to tadpoles' cancellations is 
\be{condition1}
\sum_{a}N_a(\pi_{a}+\pi_{a'})=4\pi_{\Omega}
\ee
where $a=3,1,1'$ in the present case, $\pi_{a}$ 3-cycles wrapped by ``ordinary" D6-branes and $\pi_{a'}$ 3-cycles wrapped by the ''image" D6-branes. 
Condition (\ref{condition1}) can be conveniently rewritten  
in terms of field representations 
\be{condition1a}
\forall \: a\neq a' \qquad \#F_a-\#\bar{F}_a+(N_{a}+4)(\# S_{a}-\# \bar{S}_{a})+(N_{a}-4)(\# A_{a}-\# \bar{A}_{a})=0\ee
where $F_a,\bar{F}_a,S_{a},\bar{S}_{a}, A_{a}, \bar{A}_{a}$ are fundamental, symmetric, antisymmetric
of $U(N_{a})$ and their conjugate.
For $N_{a}>1$ these coincide with the absence of irreducible $SU(N_{a})^{3}$ triangle anomalies. 
For $N_a=1$, these are stringy conditions that can be rephrased as absence of `irreducible' $U(1)^3$ {\it i.e.~} those arising from inserting all the vector bosons of the same $U(1)$ on the same boundary. 
Let us explicitly check tadpole cancellation for the $3,1,1'$ nodes:
\be{nodethree}
U(3):\,\,\,\,\,2 n_{Q}-n_{D+D'}-n_{U}-(3-4) n_{C}=6-4-3+1=0
\ee
\be{nodeuno}
U(1):\,\,\,\,\,2 n_{L}-n_{E}-n_{N}=6-3-3=0
\ee
\be{nodeunop}
U(1)':\,\,\,\,\,n_{E}-n_{N}+3 n_{D+D'}-3 n_{U}+(1-4) n_{\Xi}=3-3+3{\cdot}4-3{\cdot}3-3=0 
\ee
Notice that anti-symmetric representations of $U(1)$'s, though trivial in that they don't correspond to any physical state, 
contribute to the tadpole conditions. In particular the arrow $\Xi$, connecting the node $U(1)'$ with its image, contributes to tadpole cancellation in the node $1'$. 

The condition for a massless vector boson associated to $U(1)_{Y}$, with $Y=\sum_{a}c_{a}Q_{a}$,
\be{condition2}
\sum_{a}c_{a}N_{a}(\pi_{a}-\pi'_{a})=0
\ee
can be translated into 
 \be{condition3a}
c_{a}N_{a}(\#S_{a}-\# \bar{S}_{a}+\# A_{a}-\# \bar{A}_{a})-\sum_{b\neq a} c_{b}N_{b}(\#(F_a,\bar{F}_b) - \#(F_a,F_b))=0
 \ee
One can verify that also these conditions are satisfied for each nodes in Fig.~3 for $c^Y_3=1/{3}$, $c^Y_1=-1$, $c^{'Y}_1=1$. Once again for $U(1)'$ one can either include $\Xi$ in the counting or replace its contribution in terms of `fundamentals' using 
tadpole cancellation. The massive (anomalous) $U(1)$'s are associated to $3Q_3 + Q_1$ and to $3Q_3 - Q'_1$, as can be seen computing the anomaly polynomial. It is amusing to observe that, removing $D', C$ and $\Xi$, any combination of $B-L=Q_3/3 - Q_1$ and $T_R = Q'_1/2$ would remain massless.

We should remark that the quiver represented in Fig.~3 could not generate 
extra R-parity breaking terms $\lambda'LQD^{c}$ in (\ref{WR}) or $\lambda''' LQD'^{c}$, leading to a mixing of quarks and leptons. Due to (anomalous) gauge symmetries no renormalizable R-parity violating coupling can be generated. Dangerous higher-dimension operators, associated to polygons with more than three sides, may appear that are either suppressed or altogether absent since not all closed polygons necessarily correspond to interaction terms.

Let us note that quiver in Fig.3 is inspired by a Pati-Salam models $Sp_{L}(2)\times Sp_{R}(2) \times U(4)$:
it can be operatively obtained from a Pati-Salam-like quiver. 

As an alternative, we can consider the quiver in Fig.~5, with right-handed neutrini with $Y=0$ stretching from $U'(1)$ to $U(1)''$. The Dirac mass term $H^uNL$ is present already at the perturbative level, while the Majorana mass term $m_N N^2$ can be generated by exotic instantons, corresponding either to $O(1)$ E2-brane instantons with double intersections with both $U(1)$ and  $U(1)'$ to $U(1)''$ or to $Sp(2)$ E2-brane  instantons with single intersection. (In Fig.~5, we omit the presence of this third $E2$-instanton). 

Despite the presence of $N$, lepton number is only violated non-perturbatively and no additional R-parity breaking terms like $\mu_{a}L_{a}H_{u}$ with $\Delta L=1$ are present at the perturbative level. Yet, integrating out $N$, $H^u$, $H^d$, $C$ and $D'_c$ would lead to dangerous baryon and lepton number violating terms as discussed in Section~3.1. 

We can assume that such a Majorana mass term is not even generated non-perturbatively by turning on back-ground fluxes that would prevent E2-wrapping the relevant 3-cycle if present at all. In the limit $m_N\rightarrow 0$ only $QQQQU^{c}/\mu\mathcal{M}_{0}$ remains, 
but this cannot lead to proton decay, as discussed in \cite{Addazi:2014ila}. 

Since other R-parity violating terms in (\ref{WR}) are automatically 
disallowed at the perturbative level, our model is R-parity invariant to start with.  Taking into account the non-perturbative term $QQQH$ indirectly generated through exotic instantons, the Majorana mass of $N$
and the $\mu$-term $\mu H_{u}H_{d}$ possibly generated by exotic instantons, one can expect new higher-order terms of the form
\begin{equation}\label{op}
\mathcal{W}_{n>3}=y_{H^uLH_{d}D^{c}Q}\frac{1}{M^2_{S}}H^uLH_{d}D^{c}Q+y_{U^{c}QH_{u}H^dD^{c}}\frac{1}{M^2_{S}}H_{u}U^{c}QH_{d}D^{c}
\end{equation}
not present in the previous case.
Clearly these operators are dangerous. For instance, combining $\mathcal{W}_{n>3}$ with the non-perturbative operator (\ref{effective}) with (\ref{op}), yields 
\begin{equation}\label{danger}
\mathcal{W}'_{eff} = \frac{1}{M_{S}\mathcal{M}_{0}\mu}QQQQ^cU^{c}D^{c}+\frac{1}{M_{S}\mathcal{M}_{0}\mu}QQQLQD^{c}
\end{equation}
where $Q^c$ denotes either $U^c$ or $D^c$. The first term can lead to neutron-antineutron transitions and di-nucleon decays $pp\rightarrow \pi^{+}\pi^{+},K^{+}K^{+}$, 
the second term to proton decay $p\rightarrow \pi^{0}e^{+}$.
The ratio of the proton life-time to the neutron-antineutron transition time is
\begin{equation}\label{time}
\tau_{n\bar{n}}\simeq \frac{\mathcal{M}_{0}}{M_{S}}\tau_{p-decay}\simeq e^{-S_{E_{2}}}\tau_{p-decay}
\end{equation}
This hierarchy is much higher than the present 
limit on $D'-C$ vector-like pairs at colliders. 
In fact, with $\tau_{p-decay} \simeq 10^{34\div 35}\, \rm yr$ and
$\tau_{n\bar{n}}\simeq 3 \, \rm yr$, $\mathcal{M}_{0} \simeq 10^{-35}M_{S} <<\mathcal{M}_{0}|_{exp}$, where $\mathcal{M}_{0}|_{exp}\simeq \rm 0.5\div 1 \rm TeV$ is the direct bound from colliders discussed above. 
For di-nucleon decay the situation is better, but also in this case the required tuning is extremely delicate, considering that $\tau_{di-decay}\simeq 10^{32}\, \rm yr$ \cite{PDG}.
So, we conclude that a fine tuning of the coupling constants  
$y_{H_uLH_{d}D^{c}Q}$, $y_{H_uU^{c}Q^{c}H_{d}D^{c}}$ close to zero would be necessary
in this case. 

\begin{figure}[t]
\centerline{ \includegraphics [height=6cm,width=0.3 \columnwidth]{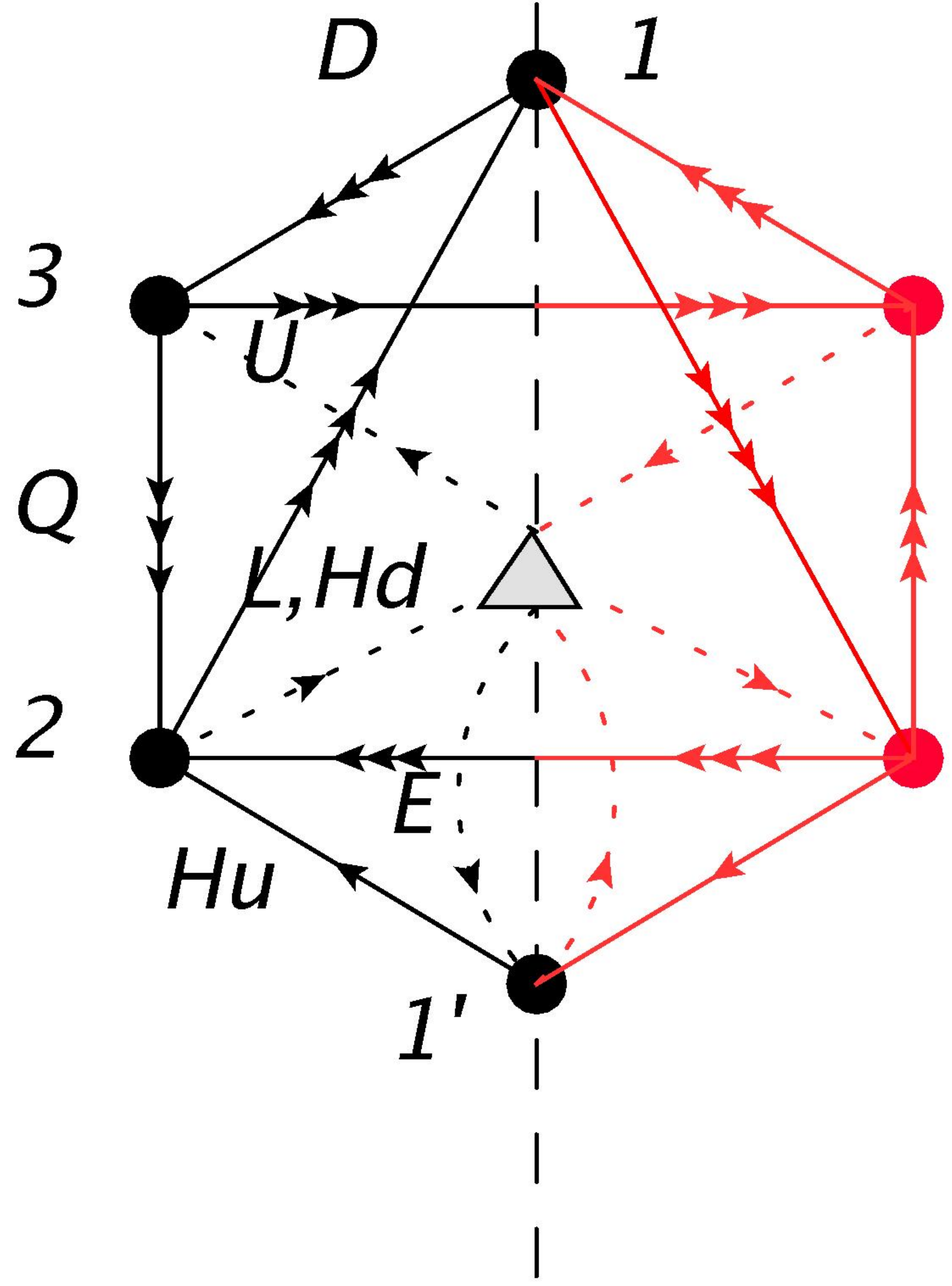}}
\vspace*{-1ex}
\caption{Another example of quiver, inspired by $U(5)$ models. In this model, not 
all standard Yukawa couplings are generated.
For simplicity, we show only the $E2$-instanton, non-perturbatively generating $H_{u}U^{c}Q$. We omit $D',C$
in this figure.  }
\label{plot}   
\end{figure}

Another class of un-oriented quivers where a Majorana mass could be generated non-perturbatively, yet in a more contrived way is based on the quiver of Fig.~4. The hyper-charge generator is $Y=2Q_3/3 - Q_2$. Many `standard' Yukawa couplings, including $H_u U^c Q$, are disallowed in this case. They can be generated non-perturbatively as in $U(5)$ D-brane models, while undesired couplings such as $LD^cQ$ should be tuned to zero. 
Right-handed neutrini can be inserted between $1-1'$ in Fig.~6.
Extra exotic matter has to be considered in order to satisfy tadpoles' cancellations and massless hypercharge conditions.
We will not discuss this class of model any further here.  

\subsection{Extended quivers and CY singularities}

The quivers, proposed in Fig.~3 or Fig.~4, have different numbers of arrows entering/exiting each node. 
As a consequence, these do neither seem to be systems of (fractional) D3-branes transverse
to a local Calabi-Yau singularity nor T-dual to these with D6-branes. 
This condition seems to be a general rule before the $\Omega$-projection of the system.
Fig.~3 and Fig.~4 can be viewed as subsystems of larger quivers with empty nodes, possibly corresponding to flavor branes or exotic instantons, that are not needed for tadpole/anomaly cancellation. Yet one can restore in this way a perfect balance of entering/exiting arrows and interpret the above quivers as projections of systems on a local orbifold or CY singularity, in the presence of $\Omega$-planes and Flavor Branes. Alternatively one can introduce /remove arrows among the nodes that correspond to (chiral) super-fields
in anomaly/tadpole free combinations that can be made very massive by Yukawa couplings, fluxes or exotic instantons.

In general, configurations, with flavor branes and orientifold planes $\Omega$, 
can provide examples of realistic models for particle physics, containing SM.
In these, the super-conformal $\mathcal{N}=4$ theory 
is broken to an $\mathcal{N}=1$ theory. 
The low-energy dynamics is governed by a local
unoriented quiver theory, in which consistency at the quantum level depends 
 on local tadpole cancellation.  For some examples of unoriented quivers with Flavor,
 it is possible to show explicitly relations between tadpoles and anomalies
 in presence of flavor branes \cite{18,19,20,BIMP}

Among the variety of possibilities, we propose a simple extension of the quiver in Fig.~3, shown in Fig.~5. We would like to stress that this is only one example among different possible quivers. 
In this case the `flavor' brane is an empty node ($N_0 = 0$) in the quiver invariant under $\Omega$. As
a consequence, one has to be careful about non-perturbative effects possibly generated by an exotic instanton in this node. In the example shown in Fig.~5 one can easily check that the putative $E2$ does not lead to any  
extra superpotential term. 
Let us also observe that the extra arrows neither contribute to 
tadpoles nor to the mass of the hypercharge vector boson. 

\begin{figure}[t]
\centerline{ \includegraphics [height=8cm,width=0.6 \columnwidth]{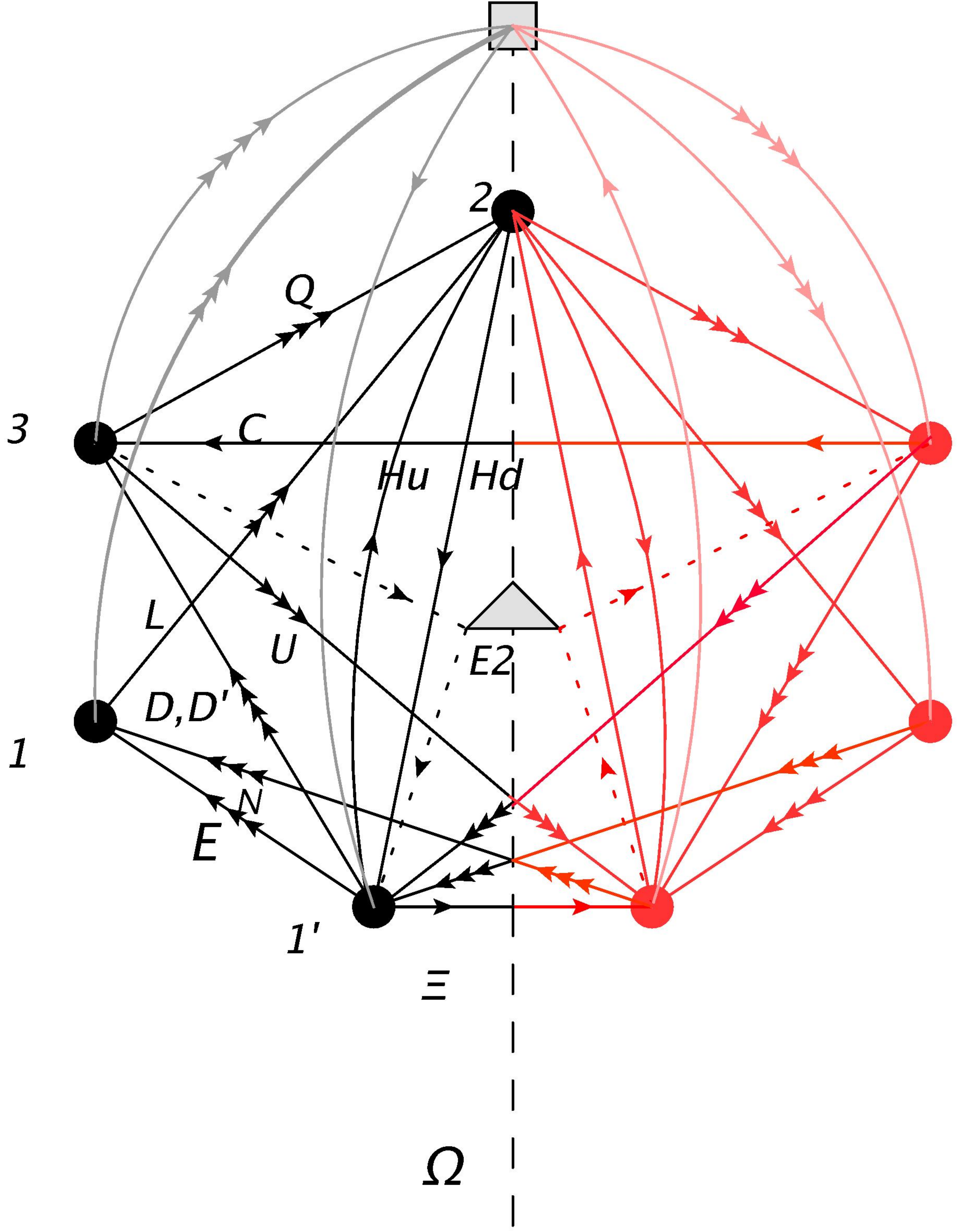}}
\vspace*{-1ex}
\caption{An example of a simple extension of the quiver in Fig.~3, with a Flavor brane (square). }
\label{plot}   
\end{figure}

\section{K\"ahler potential, D-terms and perturbative corrections}

So far we have focussed on the super-potential interactions, both 
perturbative and non-perturbative ones. We have argued that barring explicit R-parity violating 
terms in the Lagrangian, R-parity is broken dynamically by 
non-perturbative exotic instanton effects. This implies that it is 
preserved in perturbation theory, at least in so far as we keep 
supersymmetry unbroken. Since supersymmetry has to be broken by `soft 
terms' one may be worried about proton decay and other undesired effects. 
However, even before addressing the issues 
related to soft SUSY breaking, one may wonder whether D-terms and 
corrections to the K\"ahler potential may affect our analysis 
significantly. Although little is known about quantum corrections to the 
Kahler potential and D-terms in the intersecting D-brane 
models, some progress has been made in
\cite{Grana:2014vva, Berg:2014ama, Berg:2005ja, Billo:2007py}.
The main idea is to use in a sense the locally supersymmetric version of 
the exact Novikov-Shifman-Vainstein-Zakharov $\beta$ function in order 
to derive an exact (perturbative) relation between corrections to $K(\Phi, \Phi^\dagger)$ and thus  
anomalous dimensions $\gamma$, related to wave-function renormalisation $Z_\Phi$, and running of $g_{YM}$ and thus $\beta$ function. Except for theories or sectors with at least $N=2$ susy, 
whereby the K\"ahler potential for the vector multiplet is directly 
related to the holomorphic pre-potential and thus to the gauge couplings {\it i.e.} the gauge kinetic function
and can be computed, when susy is minimal {\it i.e.} $N=1$, the relation is 
much looser.
In principle $K$ and the D-terms in general, can get any sort of 
perturbative corrections. However these are to be compatible with the 
`classical' symmetries, which include R-parity, baryon number $B$ and 
Lepton number $L$. It is also known that standard `gauge' instantons
can 
only generate terms violating `anomalous' symmetries, while
`exotic' 
instanton can violate non-anomalous ones, such as $B-L$ in the (MS)SM. 
Depending on the number of fermionic zero-modes both gauge and exotic 
instantons may correct the gauge kinetic function(s), D-terms and the 
Kahler potential. It is rather reasonable to assume that such 
non-perturbative corrections be absent or very small in the quiver 
models in our classes, even when the string scale is close to -- but smaller than -- the Planck 
scale so much so that the full super-gravity structure should be taken 
into account. 

In summary, the only `seed' of R-parity breaking and Baryon (and/or 
Lepton) number violation seems to be the super-potential.

When supersymmetry gets broken, say in a hidden (strongly coupled) 
sector and then communicated to the visible sector, the situation gets 
more intricate. The structure of the low-energy Lagrangian, though 
constrained by the original supersymmetry, allows for dangerous mixings. 
In the quiver models we consider, proton stability, as previously discussed, 
largely relies on Lepton number conservation or on the fact that the 
final states should contain at least one susy partner. In Pati-Salam like models, it's built in via the selection rule $\Delta B = 2$.

\section{Conclusions}

We have produced two examples of consistent quiver fields theories, indirectly generating 
a Majorana mass term for the neutron by means of exotic instantons.
These are free of local tadpoles and thus irreducible anomalies. 

The phenomenology exposed by this class of models is interesting
both for neutron-antineutron physics and LHC or other colliders, where a new vector-like pair could be detected. 
On the other hand, the models we suggest can be tuned to suppress FCNC's. However, in order to prevent fast proton decay, Lepton number is not to be violated. An alternative is to consider $SO(10)$ GUT models or Pati-Salam like models in string theory that can lead to $\Delta B = 2$ processes but no  $\Delta B = 1$ \cite{PS0,PS1,PS2,PS3,PS4,PS5}. Although perturbative un-oriented strings do not admit spinor representations of orthogonal groups, P-S like models are easy to embed in this context \cite{Anastasopoulos:2010ca}. In $SO(10)$ neutrino Majorana masses are generated by Higgses in the ${\bf 126}$ that involve $({\bf 10, 3, 1}) + ({\bf 10^*, 1, 3})$ of the PS group 
$SU(4)\times SU(2)_L\times SU(2)_R$. These cannot appear either in perturbative open string settings. Yet
combining $({\bf 10, 1, 1}) + ({\bf 10^{*}, 1, 1})$ with $({\bf 1, 3, 1}) + ({\bf 1, 1, 3})$ that are allowed one can achieve the goal of first breaking $SU(4)$, then $SU(2)_R$ and finally $SU(2)_L \times U(1)_Y$ to $U(1)_{e-m}$ and get Majorana neutrini and neutrons with a stable proton. 
We plan to discuss this class of models in a forthcoming paper.

In principle, it is possible to construct various quivers 
with flavor branes, generating other fascinating effects for phenomenology. 
A complete classification could reserve us some surprises. 
It remains to search Calabi-Yau compactifications leading to global embeddings of models of this kind. 

To conclude, the class of models considered represents an intriguing example 
of a phenomenological effective model of string theory 
beyond the standard model, that could be tested by the next generation
of experiments.

\vspace{1cm} 

{\large \bf Acknowledgments} 
\vspace{3mm}

A.~A would like to thank Galileo Galilei Institute for Theoretical Physics 
for hospitality, where this paper was prepared. 
We would like to thank Matteo Bertolini, Francisco Morales, Fernando Quevedo, Andrea Romanino, Marco Serone, Giovanni Villadoro, Jim Halverson and Angel M.Uranga for useful comments. 
We would also like to thank the anonymous referee for important suggestions and remarks in these subjects.

\section*{Appendix: Integrating out massive super-fields}
We find it more intuitive to apply a perturbative approach that we report in the following for pedagogical purposes.

Setting $M_{F}=M^{0}_{F}+\mathcal{E}$, 
with 
\begin{equation}\label{Matrix2}
M^{0}_{F} = \left( \begin{array}{ccccc} m_{n} & 0 & 0 & 0 & 0
\ \\  0 & 0 & \mu & 0 & 0  \ \\
0 & \mu & 0 & 0 & 0 \ \\
0 & 0 & 0 & 0 & \mathcal{M}_{0} \\\
0 & 0 & 0 & \mathcal{M}_{0} & 0 
\end{array} \right)
\end{equation}
\begin{equation}\label{Matrix3}
\mathcal{E}= \left( \begin{array}{ccccc} 0 & L & 0 & 0 & 0
\ \\  L & 0 & 0 & 0 & 0  \ \\
0 & 0 & 0 & 0 & Q \ \\
0 & 0 & 0 & 0 & 0 \\\
0 & 0 & Q & 0 & 0 
\end{array} \right)
\end{equation}
the inverse mass matrix can be calculated as a perturbative series 
\begin{equation}\label{perturbationSeries}
M_{F}^{-1}=(M_{F}^{0})^{-1}-(M_{F}^{0})^{-1}\mathcal{E}(M_{F}^{0})^{-1}
\end{equation}
$$+(M_{F}^{0})^{-1}\mathcal{E}(M_{F}^{0})^{-1}\mathcal{E}(M_{F}^{0})^{-1}+(M_{F}^{0})^{-1}\mathcal{E}(M_{F}^{0})^{-1}\mathcal{E}(M_{F}^{0})^{-1}\mathcal{E}(M_{F}^{0})^{-1}+...$$
In our case, combining (\ref{Matrix3}) and the inverse of $(\ref{Matrix2})$
one gets the first perturbation  
\begin{equation}\label{Matrix4}
(M_{F}^{0})^{-1}\mathcal{E}(M_{F}^{0})^{-1}= \left( \begin{array}{ccccc} 0 & 0 & \frac{L}{m_{N}\mu} & 0 & 0
\ \\  0 & 0 & 0 & 0 & \frac{Q}{\mu \mathcal{M}_{0}}  \ \\
\frac{L}{\mu \mathcal{M}_{0}} & 0 & 0 & 0 & 0 \ \\
0 & \frac{Q}{\mu \mathcal{M}_{0}} & 0 & 0 & 0 \\\
0 & 0 & 0 & 0 & 0 
\end{array} \right)
\end{equation}
then the second perturbation is 
\begin{equation}\label{Matrix5}
(M_{F}^{0})^{-1}\mathcal{E}(M_{F}^{0})^{-1}\mathcal{E}(M_{F}^{0})^{-1}= \left( \begin{array}{ccccc} 0 & 0 & 0  & \frac{LQ}{m_{N}\mu\mathcal{M}_{0}} & 0 
\ \\  0 & 0 & 0 & 0 & 0  \ \\
0 & 0 & \frac{L^{2}}{m_{N}\mu^{2}} & 0 & 0 \ \\
\frac{QL}{m_{N}\mu \mathcal{M}_{0}} & 0 & 0 & 0 & 0 \\\
0 & 0 & 0 & 0 & 0 
\end{array} \right)
\end{equation}
the third perturbation is 
\begin{equation}\label{Matrix6}
(M_{F}^{0})^{-1}\mathcal{E}(M_{F}^{0})^{-1}\mathcal{E}(M_{F}^{0})^{-1}\mathcal{E}(M_{F}^{0})^{-1}= \left( \begin{array}{ccccc} 0 & 0 & 0  & 0 & 0 
\ \\  0 & 0 & 0 & 0 & 0  \ \\
0 & 0 & 0 & \frac{L^{2}Q}{m_{N}\mu^{2}\mathcal{M}_{0}} & 0 \ \\
0 & 0 & \frac{L^{2}Q}{m_{N}\mu^{2}\mathcal{M}_{0}} & 0 & 0 \\\
0 & 0 & 0 & 0 & 0 
\end{array} \right)
\end{equation}

At the fourth order, we recover the exact result cited above in the paper. 


\end{document}